\documentclass[twocolumn,a4paper,english,a4paper,preprintnumbers,amsmath,amssymb,prb]{revtex4}
\usepackage[latin1]{inputenc}
\usepackage{float}
\usepackage{graphicx}
\usepackage{amssymb}

\makeatletter

\usepackage{float}
\makeatletter

\usepackage{dcolumn}

\usepackage{bm}

\makeatother

\usepackage{babel}
\makeatother
\begin{document}

\title{Interplay between Hund's coupling and spin-orbit interaction 
on elementary excitations in Sr$_2$IrO$_4$}

\author{Jun-ichi Igarashi$^{1}$ and and Tatsuya Nagao$^{2}$}

\affiliation{
$^{1}$Faculty of Science, Ibaraki University, Mito, Ibaraki 310-8512,
Japan\\
$^{2}$Faculty of Engineering, Gunma University, Kiryu, Gunma 376-8515,
Japan
}

\date{\today}

\begin{abstract}
We study the elementary excitations in 5d transition metal oxide
Sr$_2$IrO$_4$ by calculating the particle-hole Green's function within the random phase approximation on an antiferromagnetic ground state in the two-dimensional multi-orbital Hubbard model. The obtained magnetic excitations of bound states show a characteristic dispersion in consistent with the experiments. In addition, two new types of excitations are found due to the interplay between spin-orbit interaction and Hund's coupling: a magnetic excitation as a bound state, which has energy gap at the $\Gamma$ point, and an exciton as a resonant mode in the continuum of electron-hole pair creation.
\end{abstract}

\pacs{71.10.Fd 75.25.Dk 71.10.Li 71.20.Be}

\maketitle
Electron correlation effects on transition metals and their compounds
have attracted much interest since the discovery of high-$T_{\textrm{C}}$ cuprate superconductors.  In the $3d$ transition metal ions, the spin orbit interaction (SOI) plays a minor role on the electronic structures, since it is usually much smaller than the on-site Coulomb interaction and the crystal field splitting. In the $5d$ transition metal ions, however, the SOI is one order of magnitude larger than that of the $3d$ systems while the Coulomb interaction becomes weaker due to the 
extended nature of the $5d$ electrons.  Accordingly, the interplay between the 
electron correlation and the SOI is expected to bring about new intriguing phenomena. 
For this reason, much attention has recently been paid to Ir oxides such as Sr$_2$IrO$_4$\cite{Crawford1994,Cao1998,Moon2006,Clancy2012,Haskel2012,Ye2013,Carter2013,Watanabe2013} and 
Na$_2$IrO$_3$.\cite{Singh2010,Singh2012,Chaloupka2013,Trousselet2013}

In particular, we focus on Sr$_2$IrO$_4$, which consists of two-dimensional IrO$_2$ layers showing structural similarity to the parent compound of high-$T_{\textrm{C}}$ cuprate La$_2$CuO$_4$, and exhibits a canted antiferromagnetic (AFM) order below $230$ K.\cite{Crawford1994,Cao1998,Moon2006}
The energy of the $e_g$ orbitals is estimated about 2 eV higher than 
that of the $t_{2g}$  orbitals due to the large crystal field.
Five $5d$ electrons are occupied per Ir atom, and one hole is sitting in the t$_{2g}$ orbitals.
Since the hole states have an effective orbital angular momentum 
$\ell$ equal to $-1$, the lowest-energy states on the localized 
electron picture are doubly degenerate with the effective total 
angular momentum $j_{\rm eff}=\frac{1}{2}$ under the SOI:\cite{Kim2008,Kim2009} 
$|\phi_{\pm}\rangle=\frac{1}{\sqrt{3}}\left(|yz,\mp\sigma\rangle
\pm i|zx,\mp\sigma\rangle \pm|xy,\pm\sigma\rangle \right)$,
where $yz$, $zx$, and $xy$ designate $t_{2g}$ orbitals, and spin component
$\sigma=\uparrow$.
By introducing the isospin operators acting on these states, the system is mapped onto the Heisenberg model, from which an insulating AFM phase is derived, consistent with the experiments.\cite{Crawford1994,Cao1998,Moon2006}
Furthermore, it has been pointed out that the small anisotropic 
terms emerge in addition to the isotropic term, when Hund's coupling 
is taken into account in the second-order process in the strong coupling
expansion.\cite{Jackeli2009,Kim2012}

Recently, resonant inelastic x-ray scattering (RIXS) experiment at the 
Ir $L$ edge has detected the excitation spectra,\cite{J.Kim2012}
whose low energy part 
follows the dispersion relation similar to the spin wave in the Heisenberg
model on a square lattice. A notable point is that
the excitation energy at the M point is nearly 
half of that at the X point,
in contrast to the situation in the undoped cuprates such as 
La$_2$CuO$_4$\cite{Braicovich2009}
and Sr$_2$CuO$_2$Cl$_2$.\cite{Guarise2010}
It is known that the Heisenberg model with only the nearest neighbor exchange interaction
gives the spin-wave energy at the M point nearly the same as 
that at the X point.
The large energy difference in Sr$_2$IrO$_4$ requires the large farther 
neighbor exchange interactions in the Heisenberg model, 
which suggests the importance of itinerant character. 
The excitonic excitations are observed 
around 0.4-0.9 eV, not far above the magnetic excitations.\cite{J.Kim2012}
This contrasts with the excitation spectra of undoped cuprates, where
only the d-d excitations appear with the energies far above the spin wave
excitations.\cite{Braicovich2009,Guarise2010} 
In addition to these experimental facts,
a study using the dynamical mean field theory has argued
that the system is not the Mott insulator but the Slater 
insulator.\cite{Arita2012} In contrast, a recent angle-resolved
photoemission measurement suggests that the data are consistent
with a Mott scenario rather than a Slater scenario.\cite{Moser2014}

Under such circumstances, it may make sense to study elementary excitations
from the intermediate coupling scheme based on the itinerant electron picture.
In this paper, introducing the multi-orbital Hubbard model to describe
the system, we calculate the particle-hole Green's function
within the Hartree-Fock approximation (HFA) 
and the random phase approximation (RPA).
We first confirm the obtained magnetic excitations as bound states  
show good agreement with the RIXS experiment.\cite{J.Kim2012}
Furthermore, we find two kinds of new excitation modes emerge, 
which are attributed to the interplay between the SOI and Hund's 
coupling. 
One is the gap mode associated with 
the splitting of the magnetic excitation. 
Another is the exciton in the continuum of electron-hole
pair excitation, whose dispersive behavior as a function of
the momentum transfer shows in qualitative agreement with the RIXS experiment.  

We employ the multi-orbital Hubbard model defined by
the base states in the local coordinate frames rotated 
in accordance with the rotation of the oxygen octahedra 
surrounding an Ir atom with respect to the crystallographic 
$c$ axis about 11$^{\circ}$.\cite{Watanabe2010,Wang2011}
It may be expressed as
\begin{equation}
 H =  H_{\rm kin}+H_{\rm SO}+H_{\rm I},
\end{equation}
with
\begin{eqnarray}
H_{\rm kin} & = & \sum_{\left\langle i,i'\right\rangle }
\sum_{n,n',\sigma}\left(t_{in,i'n'}d_{in\sigma}^{\dagger}d_{i'n'\sigma}+ {\rm H.c.}
\right),\\
H_{\rm SO} & = & \zeta_{\rm SO}\sum_{i}
\sum_{n,n',\sigma,\sigma'}
d_{in\sigma}^{\dagger}({\bf L})_{nn'}
 \cdot({\bf S})_{\sigma\sigma'}d_{in'\sigma'}, \\
H_{\rm I} & = & 
 U\sum_{i,n} n_{in\uparrow}n_{in\downarrow} \nonumber \\
 \nonumber\\
&+&J\sum_{i,n\neq n'} (d_{in\uparrow}^{\dagger}d_{in'\downarrow}^{\dagger}
                     d_{in\downarrow}d_{in'\uparrow}
                    +d_{in\uparrow}^{\dagger}d_{in\downarrow}^{\dagger}
                     d_{in'\downarrow}d_{in'\uparrow}) \nonumber \\
&+&\sum_{i,n<n'\sigma}[U' n_{in\sigma}n_{in'-\sigma}
                 + (U'-J) n_{in\sigma}n_{in'\sigma}],
\end{eqnarray}
where $d_{in\sigma}$ denotes the annihilation operator of 
an electron with orbital $n$ ($=yz,zx,xy$) and spin $\sigma$ at the Ir site $i$.
The $H_{\rm kin}$ represents the kinetic energy with transfer integral 
$t_{in,i'n'}$. An electron on the $xy$ orbital could transfer to the $xy$ 
orbital in the nearest neighbor sites through the intervening O $2p$ orbitals,
while an electron on the $yz$($zx$) orbital could transfer to 
the $yz$($zx$) orbital in the nearest-neighbor sites only along the $y$($x$) 
direction.
The none-zero values of $t_{in,i'n'}$'s are assumed to be the 
same and denoted as $t_1$.
The crystal distortion makes the energy of the $xy$ orbital different from 
that of the $yz$ and $zx$ orbitals, as well as the further neighbor transfer
integrals with the $xy$ orbital substantial, which effects are neglected.
The $H_{\rm SO}$ represents the SOI of $5d$ electrons; 
$({\bf L})_{nn'}$ represents the matrix element of the orbital momentum
operator between the orbitals specified by $n$, $n'$, 
and $({\bf S})_{\sigma\sigma'}$ represents the matrix element of spin 
angular momentum operators between the spin states specified by
$\sigma$, $\sigma'$.
The $H_{\rm I}$ represents the Coulomb interaction between electrons,
which satisfies $U=U'+2J$.\cite{Kanamori1963}
We use the values $U=1.26$ eV, $\zeta_{\rm SO}=0.324$ eV, $t_1=0.324$ eV,
and $J/U=0-0.15$ in the following calculation. 

We consider a unit cell $j$ containing two atoms at ${\bf r}_i(j)=
\textbf{r}_j +\mbox{\boldmath{$\delta$}}_i$, where 
$\mbox{\boldmath{$\delta$}}_i=(0,0)$ and $(a,0)$ for
sublattices A and B, respectively. Here $a$ is a nearest neighbor 
distance. 
The Fourier transform of annihilation operator is defined in the half of the
Brillouin zone called as the magnetic Brillouin zone (MBZ):
\begin{equation}
d_{\lambda n\sigma}({\bf k}) = \sqrt{\frac{2}{N}}\sum_{j}d_{in\sigma}
                             e^{-i{\bf k}\cdot{\bf r}_j},
\end{equation}
where $j$ runs over $N/2$ unit cells and $\textbf{k}$ is the wave number.
The sublattice A or B is discriminated by $\lambda=1$ or 2, respectively. With this definition together with abbreviations $\xi=(\lambda,n,\sigma)$ 
and $\xi'=(\lambda',n',\sigma')$, $H_{\rm kin}$ is rewritten as
\begin{equation}
 H_{\rm kin} = \sum_{{\bf k}\xi\xi'} d_{\xi}^{\dagger}({\bf k})
             \left[\hat{H}_{\rm kin}({\bf k})\right]_{\xi,\xi'}d_{\xi'}({\bf k}).
\end{equation}
Then, we introduce the single-particle Green's function 
in a matrix form with $12\times 12$ dimensions,
\begin{equation}
\left[\hat{G}({\bf k},\omega)\right]_{\xi,\xi'}=-i\int\langle 
 T(d_{\xi}({\bf k},t)d_{\xi'}^{\dagger}({\bf k},0))\rangle
 e^{i\omega t}{\rm d}t,
\label{eq.dG}
\end{equation}
where $T$ is the time ordering operator, and $\langle X \rangle$
denotes the ground-state average of operator $X$.

We follow the conventional procedure of the HFA with the help of
the Green's function. The summation over ${\bf k}$ is
carried out by dividing the MBZ into
$100\times 100$ meshes to evaluate average values of density operators.\cite{Igarashi2013-1}
Assuming the staggered moment along the x-axis,
we obtain a self-consistent solution of the AFM order 
consistent with the magnetic measurements.
\cite{Crawford1994,Cao1998}
Both the orbital and spin moments are induced due to the strong 
SOI. For $J/U=0.15$, we have 
$\langle S_x\rangle= \pm 0.112$ and 
$\langle L_x\rangle= \pm 0.435$.
These values are compared with the average on the Kramers' doublet 
$|\frac{1}{2},\pm\frac{1}{2}\rangle\equiv \frac{1}{\sqrt{2}}
(|\phi_{\pm}\rangle\pm|\phi_{\mp}\rangle)$ defined 
with the quantization axis along the $x$ axis, that is:
$\langle\frac{1}{2},\pm\frac{1}{2}|S_x|\frac{1}{2},\pm\frac{1}{2}
\rangle= \mp \frac{1}{6}$ and 
$\langle\frac{1}{2},\pm\frac{1}{2}|L_x|\frac{1}{2},\pm\frac{1}{2}
\rangle= \mp \frac{2}{3}$.
Note that the obtained AFM order in the local coordinate frames 
implies that the canted AFM order is realized in the global 
coordinate frame with concomitant weak ferromagnetic component.

\begin{figure}
\includegraphics[width=8.0cm]{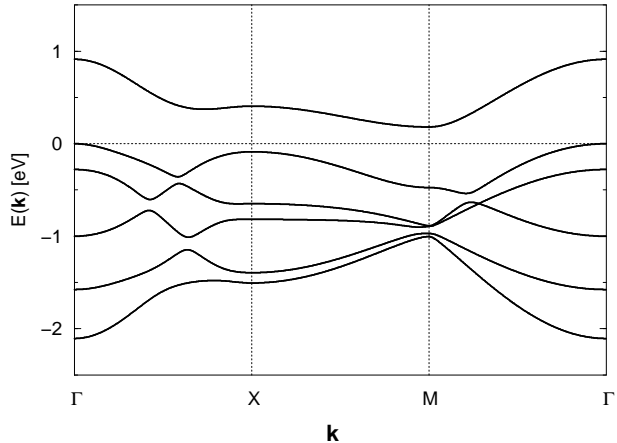}
\caption{\label{fig.disp} Single-particle energy along symmetry lines.
Here, the special points $\Gamma$, $X$, and $M$ refer to 
$\textbf{k}=(0,0)$, $(\pi,0)$, and 
$\left(\frac{\pi}{2},\frac{\pi}{2}\right)$, respectively. 
$J/U=0.15$. The origin of energy is set at the top of valence band.}
\end{figure}

The single-particle Green's function is expressed as 
\begin{equation}
\hat{G}({\bf k},\omega)=\hat{U}({\bf k})\hat{D}({\bf k},\omega)
\hat{U}({\bf k})^{-1},
\end{equation}
with 
\begin{equation}
[\hat{D}({\bf k},\omega)]_{j,j'}=\delta_{j,j'}\left\{\omega-E_{j}({\bf k})
+i\eta{\rm sgn}[E_{j}({\bf k})]\right\}^{-1},
\end{equation}
where ${\rm sgn}[A]$ stands for a sign of quantity $A$ and $\eta$ 
denotes a positive convergent factor.
The $E_{j}({\bf k})$'s represent the energy eigenvalues within the HFA
measured from the chemical potential.
Figure \ref{fig.disp} shows the single-particle energy as a function of $\textbf{k}$ along the symmetry lines. Each level is doubly degenerate 
in the MBZ. 
The conduction band mainly consists of the $j_{\rm eff}=\frac{1}{2}$ character,
and the energy gap is created by the AFM order, 
consistent with the previous studies.\cite{Kim2008,Kim2009,Watanabe2010}

Now we consider the particle-hole Green's function defined by 
\begin{equation}
\left[\hat{Y}^{{\rm T}}(q)\right]_{\xi_1\xi'_{1};\xi\xi'}=-i\int \langle 
T[(\rho_{{\bf q}\xi_1\xi'_{1}}(t))^{\dagger}
\rho_{{\bf q}\xi\xi'}(0)]\rangle
e^{iq_{0}t}{\rm d}t,
\label{eq.green_time}
\end{equation}
with 
\begin{equation}
  \rho_{{\bf q}\xi\xi'} = \sqrt{\frac{2}{N}}\sum_{\bf k}
   d_{\xi}^{\dagger}({\bf k+q})d_{\xi'}({\bf k}),
\end{equation}
where an abbreviation $q=(\textbf{q},q_0)$ is introduced 
for the energy $q_0$ and wave number $\textbf{q}$.
When ${\bf k+q}$ lies outside the MBZ, it is implicitly reduced back to 
the MBZ by a reciprocal lattice vector. 
The $\hat{Y}(q)$ is a matrix with $144\times 144$ dimensions.
Collecting up the ladder diagrams within the RPA,
we obtain
\begin{equation}
\hat{Y}^{{\rm T}}(q) = \hat{F}(q)[\hat{I}+\hat{\Gamma}\hat{F}(q)]^{-1} 
                     = \left[\hat{F}(q)^{-1}+\hat{\Gamma}\right]^{-1},
\label{eq.ladder}
\end{equation}
where $\hat{\Gamma}$ represents the antisymmetric vertex, 
function\cite{Com1}
$[\hat{\Gamma}]_{\xi_{2}\xi'_{2};\xi_{1}\xi'_{1}}=
\Gamma^{(0)}(\xi_{2}\xi'_{1};\xi_{1}\xi'_{2})$.
Function $\hat{F}(q)$ is given by
\begin{eqnarray}
&& [\hat{F}({\bf q},q_0)]_{\xi_2\xi'_{2};\xi_1\xi'_{1}} 
\nonumber \\
&\equiv&
-i\frac{2}{N}\sum_{{\bf k}}\int\frac{{\rm d}k_{0}}{2\pi}
[\hat{G}({\bf k+q},k_{0}+q_0)]_{\xi_{2},\xi_{1}}
[\hat{G}({\bf k},k_{0})]_{\xi'_{1},\xi'_{2}}
\nonumber \\
 & = & \frac{2}{N}
\sum_{\textbf{k},j,\ell}
 U_{\xi_{2}j}({\bf k+q})U_{\xi_{1}j}^{*}({\bf k+q})
 U_{\xi'_{1}\ell}({\bf k})U_{\xi'_{2}\ell}^{*}({\bf k})
\nonumber \\
 & \times & 
\left[\frac{[1-n_{j}({\bf k+q})]n_{\ell}({\bf k})}
 {q_0-E_{j}({\bf k+q})+E_{\ell}({\bf k})+i\eta}
\right. \nonumber \\
& & \left. 
 -\frac{n_{j}({\bf k+q})[1-n_{\ell}({\bf k})]}
 {q_0-E_{j}({\bf k+q})+E_{\ell}({\bf k})-i\eta}\right],
 \label{eq.green_pair}
\end{eqnarray}
where $n_{\ell}({\bf k})$ denotes the occupation number of the eigenstate
with energy $E_{\ell}({\bf k})$ given by the HFA.
Using a relation $1/(\omega-E\pm i\eta)=P\{1/(\omega-E)\}\mp 
i \pi \delta(\omega-E)$ in the last line of Eq.~(\ref{eq.green_pair}),
we can express $\hat{F}(q)$ as
$\hat{F}(q)=\hat{F}_{1}(q)+i\hat{F}_{2}(q)$
with $\hat{F}_{1}(q)$ and $\hat{F}_{2}(q)$ being Hermitian matrices.

First, we search for the magnetic excitations below the continuous electron-hole
creation. In evaluating Eq.~(\ref{eq.green_pair}), we sum over ${\bf k}$ 
by dividing the MBZ into $300\times 300$ meshes.
Since $\hat{F}_{2}(q)=0$ there, they come out as bound states. 
First we discuss on the $\Gamma$-point. Since $\hat{F}_1(0,q_0)$ is found
to have zero eigenvalues, $\hat{F}_1(0,q_0)^{-1}$ does not exist.
Hence, we determine the bound state by the divergent condition
for Eq.(\ref{eq.ladder}).
We find that one eigenvalue of Eq.~(\ref{eq.ladder}) 
goes to as large as $10^4$ in units of (eV)$^{-1}$ with $q_0\to 0$. 
It is taken as the divergence within the numerical errors.
This mode with zero excitation energy may be a Goldstone mode. 
In addition, we find another divergence occurs at $q_0=0.057$ eV,
indicating that two modes exist. 
For general values of ${\bf q}$, the bound states are determined 
by adjusting $q_0$
to give zero eigenvalue in $\hat{F}_1(q)^{-1}+\hat{\Gamma}$.
We find two modes exist in the entire MBZ. 
Note that, if Hund's coupling $J$ is zero, the lowest eigenvalue is doubly
degenerate, implying the absence of the split of modes.
The wavefunction of the magnetic excitation is given by the eigenfunction for
the zero eigenvalue of $\hat{F}(q)^{-1}+\hat{\Gamma}$, which is composed of a
direct product of the particle and hole states. The particle sectors are mainly
$|\frac{1}{2},+\frac{1}{2}\rangle$ and
$|\frac{1}{2},-\frac{1}{2}\rangle$ 
with $\lambda=1$ and $2$, respectively, reflecting
the character of the conduction band.

\begin{figure} 
\includegraphics[width=8.0cm]{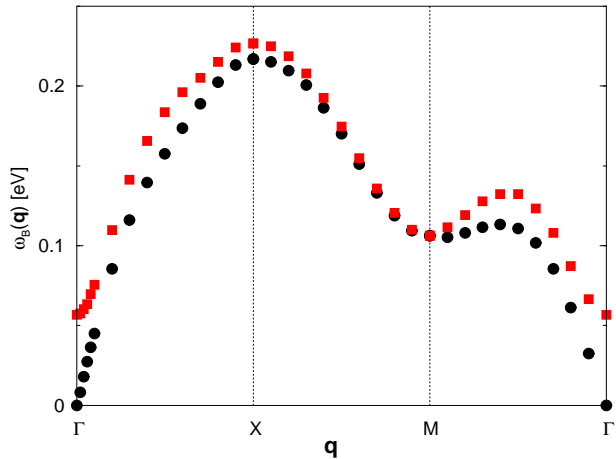} 
\caption{\label{fig.disp.mag} (Color online). Magnetic excitation energy for $\textbf{q}$ along the symmetry lines.  
$J/U=0.15$.}
\end{figure}

Figure \ref{fig.disp.mag} shows the excitation energy $\omega_B({\bf q})$
for ${\bf q}$ along high symmetry directions. 
The (black) circles and (red) 
squares represent the Goldstone and gap modes, respectively.
At the M point, the two modes take the same value of the excitation 
energy 0.105 eV, while, at the X point, they 
take the energies nearly twice 
of that at the M point, which is in good accordance with the experiment.
Notice that, within the analysis of the Heisenberg model 
on the basis of the localized picture, such result has been 
reproduced only when the second and third
nearest-neighbor exchange terms were included in addition
to the first nearest-neighbor exchange term.\cite{J.Kim2012,Igarashi2013-2}
In the present treatment, however, 
since the hopping term between the nearest neighbor sites alone 
provides the desired results, the mechanism should be sought for
different direction. It might be attributed to the 
mixing with high energy states.
Examining the wavefunction at the X point, for example, 
we actually find that the hole sectors 
$|j_{\rm eff}=\frac{3}{2},+\frac{3}{2}\rangle$ 
and $|j_{\rm eff}=\frac{3}{2},-\frac{3}{2}\rangle$
with $\lambda=1$ and $2$, respectively, have considerable amplitudes. 

The dispersion curve agrees well with the recent RIXS experiment on the whole.\cite{J.Kim2012}
It is remarkable that the RPA provides a good 
description of magnetic excitations without introducing ad hoc extended 
couplings in the Heisenberg model. 
Another remarkable point is the emergence of the gap mode.
The present authors have recently analyzed the magnetic excitations within the 
strong coupling theory, and have predicted that the gap mode is brought
about by the anisotropic exchange couplings.\cite{Igarashi2013-2}
Since the origin of the anisotropic exchange couplings is attributed to
the interplay between Hund's coupling and the SOI, the gap modes in 
the itinerant electron description are considered to have the same origin.
Although the existence of the gap mode has not been confirmed by 
experiments,\cite{J.Kim2012,Ament2011}
the RIXS experiment seems promising with further improvement 
of energy resolution enough to detect the gap.\cite{MorettiSala2013}

For $\omega$ around $\omega_B({\bf q})$, $\hat{Y}(q)$ may be expressed as
$\hat{Y}^{T}(q)=\hat{C}({\bf q})/(\omega-\omega_B({\bf q})+i\delta)$. Then, the spectral function is given by
\begin{equation}
 -2{\rm Im}\sum_{\xi\xi'}[\hat{Y}^{T}(q)]_{\xi\xi',\xi\xi'}
 =2\pi\sum_{\xi\xi'}[\hat{C}({\bf q})]_{\xi\xi',\xi\xi'}
    \delta(\omega-\omega_B({\bf q})).
\label{eq.ebspec}
\end{equation}
We evaluate numerically the weight of the pole in Eq.~(\ref{eq.ebspec}), 
and obtain the integrated intensity $I_B$ as $I_B=2\pi\times 1.20$ at the X 
point, and $2\pi\times 0.96$ at the M point with summing up the intensities 
of two modes.
Next, the spectral distribution of electron-hole pair creation is proportional
to $-2{\rm Im}\sum_{\xi\xi'}[\hat{Y}^{T}(q)]_{\xi\xi';\xi\xi'}$.
In evaluating this quantity, each $E_j({\bf k+q})-E_{\ell}({\bf k})$ 
inside the energy continuum in Eq.~(\ref{eq.green_pair}) is sorted into 
segments with the width of $0.005$ eV for $300\times 300$ ${\bf k}$-points,
resulting in the histogram representation.
Setting $q_0$ at the center of each segment, we evaluate 
Eq.~(\ref{eq.green_pair}) and thereby Eq.~(\ref{eq.ladder}).

\begin{figure}
\includegraphics[width=8.0cm]{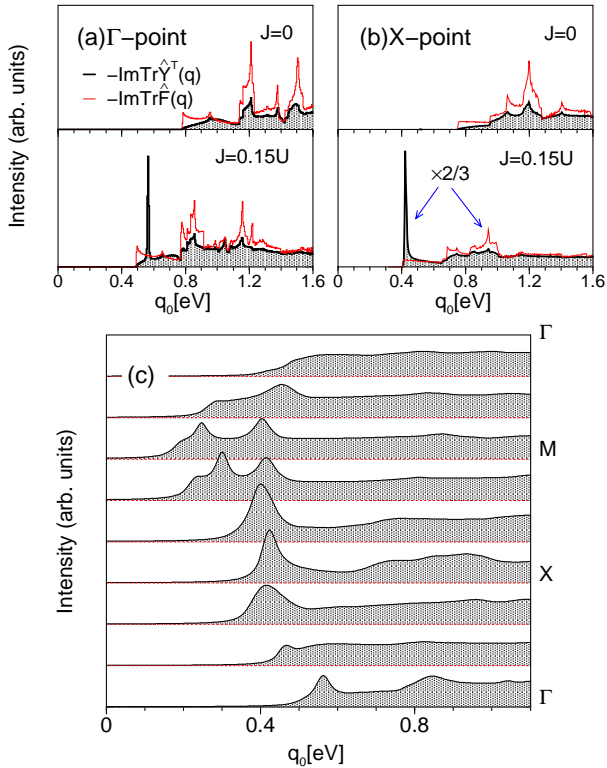} 
\caption{\label{fig.combined}
(Color online).
The thick (black) line represents the spectral function 
$-{\rm Im}$$\sum_{\xi\xi'}$$[\hat{Y}^{T}(q)]_{\xi\xi';\xi\xi'}$
of electron-hole pair creation in a fine scale.
(a) At the $\Gamma$ point and (b) X point.
$J/U=0$ (upper) and $0.15$ (lower). The thin (red) line represents
$-{\rm Im}$$\sum_{\xi\xi'}$$[\hat{F}(q)]_{\xi\xi';\xi\xi'}$.
(c) Spectral function as a function of $q_0$ with $J/U=0.15$,
which is convoluted with the Lorentzian function with 
the FWHM $0.04$ eV.
}
\end{figure}

Figures \ref{fig.combined}(a) and (b) show the spectral functions 
evaluated at the $\Gamma$ and $X$ points, respectively, 
with and without Hund's coupling.  When the multiple scattering is neglected, 
$\hat{Y}^{T}(q)$ is reduced to $\hat{F}(q)$. Then, the spectral function is given by $-2{\rm Im}\sum_{\xi\xi'}[\hat{F}(q)]_{\xi\xi';\xi\xi'}$, which is shown by the thin lines (red). Its total intensity may be expressed as 
\begin{equation}
 I_{\rm e-h}=-2{\rm Im}\int_0^{\infty}\sum_{\xi\xi'}
   [\hat{F}({\bf q},\omega)]_{\xi\xi';\xi\xi'}{\rm d}\omega.
\end{equation}
Since ten states are occupied and two states are unoccupied per unit cell,
we have $I_{\rm e-h}=2\pi\times 20$ with no $\textbf{q}$-dependence.
%, which value is independent of ${\bf q}$.

When the multiple scattering is taken into account, the spectral weight
is transferred to the lower energy region, leading to 
the decrease of intensity around $q_0\sim 1$ eV as well as the split-off of 
intensity to the bound states. In addition, a new peak, which might be called as an exciton peak, surprisingly emerges as a resonant mode in the low energy region when Hund's coupling works. 
The integrated intensities around the peaks are estimated as
$I_{\rm ex}\sim 2\pi\times 1.25$ at the $\Gamma$ point, $2\pi\times 2.12$ 
at the $X$ point, and $2\pi\times 0.6$ and $2\pi\times 1.0$ for the two
peaks at the $M$ point.\cite{Com2}
Therefore the intensities of exciton peaks
are the same order of magnitude as those of magnon peaks.

To search for the origin of the resonant mode, we examine the eigenvalues
of $\hat{F}_{1}(q)^{-1}+\hat{\Gamma}$ at the peak energy with neglecting
the small imaginary part $\hat{F}_{2}(q)$. We find that a couple of 
eigenvalues are quite close to zero, which we assign approximately as 
resonant modes. Two modes are obtained as resonant 
modes with a nearly degenerate energy at the $X$ point, while the
two modes are well separate with forming two peaks at the $M$ point.
At the first sight. the amplitudes of the corresponding eigenstate are 
distributed on many base states, but by rewriting the base states in terms 
of eigenstates with $j_{\rm eff}=\frac{1}{2}$ and $\frac{3}{2}$, 
we find at both points that the amplitudes are relatively large on the 
base states
$|\frac{1}{2},\frac{1}{2}\rangle|\frac{3}{2},-\frac{3}{2}\rangle$ and 
$|\frac{1}{2},\frac{1}{2}\rangle|\frac{3}{2},+\frac{1}{2}\rangle$ 
with $\lambda=1$, and
$|\frac{1}{2},-\frac{1}{2}\rangle|\frac{3}{2},+\frac{3}{2}\rangle$ and 
$|\frac{1}{2},-\frac{1}{2}\rangle|\frac{3}{2},-\frac{1}{2}\rangle$ 
with $\lambda=2$, where the front ket represents the excited-electron state
and the rear ket does the hole state. 
The signs of the amplitudes for $\lambda=1$ relative to those for $\lambda=2$ 
are opposite between the two modes.
Such exciton eigenstates contrast with the wavefunction of magnons,
in which the hole sectors involve
$|\frac{3}{2},+\frac{3}{2}\rangle$ and
$|\frac{3}{2},-\frac{3}{2}\rangle$
with $\lambda=1$ and $2$, respectively.
Unfortunately, the role of Hund's coupling on leading
to such eigenstates is not clear.

Figure \ref{fig.combined} (c) shows the spectral function as a function of $q_0$ along high symmetry 
directions with $J/U=0.15$, which is convoluted with the Lorentzian 
function with the FWHM $0.04$ eV.
The exciton peak moves to lower energy region with changing ${\bf q}$
from $\Gamma$ to $X$ as well as from $\Gamma$ to $M$.
This behavior is consistent with the observation in the RIXS experiment,
\cite{J.Kim2012} 
although the spectra below $q_0<0.4$ eV around the M point have not been 
detected. It should be noted that the correlation function in the 
present definition may contain spectral intensities irrelevant to the
RIXS spectra, since it has rather large integral intensity according to 
the sum-rule. For this reason, more quantitative analysis may be
necessary, which is beyond the scope of the present study,
since the RIXS spectra are not simply proportional to
$-{\rm Im}\sum_{\xi\xi'}[Y^{T}(q)]_{\xi\xi';\xi\xi'}$.
\cite{Igarashi2013-1}
In the localized electron picture, the exciton peak is interpreted 
as the excitation from the $j_{\rm eff}=\frac{3}{2}$ manifold 
to the $j_{\rm eff}=\frac{1}{2}$ manifold, which requires the energy 
$\sim\frac{3}{2}\zeta_{\rm SO}$ ($\sim 0.5$ eV), and the dispersion 
as the hopping in the AFM isospin background.\cite{Kim2012,Ament2011} 
The $j_{\rm eff}=\frac{3}{2}$ manifold 
in the present calculation forms broad bands with the width of $\sim 1.5$ 
eV, as shown in Fig. \ref{fig.disp}. 
The large amplitudes on the local excitation of $j_{\rm eff}=\frac{3}{2}
\to \frac{1}{2}$
found in the above analysis of eigenstates for excitons may partly
correspond to the localized electron picture.

In summary, we have studied the elementary excitations in Sr$_2$IrO$_4$ on
the viewpoint of itinerant electron picture.
Introducing the multi-orbital Hubbard model, we have calculated 
the particle-hole Green's function within the HFA and RPA.
We have obtained magnetic excitations as bound states  
with the dispersion relation in good accordance 
with the RIXS experiment. 
In addition, we have found that two new types of modes emerge 
due to the interplay between the SOI and Hund's coupling. One is 
the gap mode in the magnetic excitation, which is consistent with
the prediction based on the localized electron 
picture.\cite{Igarashi2013-2}
Another is the exciton in the continuum of electron-hole
pair excitation, which qualitatively captures a characteristic dependence on ${\bf q}$ shown by the RIXS experiment.
A next logical step will be to investigate the RIXS spectrum itself
since it differs from the correlation function. Such a study has been carried out by the theory based on the localized spin picture, giving a qualitative agreement with the experiment.\cite{Igarashi2014-1}
It is remarkable that a simple theory using the HFA and RPA provides
a coherent description of elementary excitations comparable to 
experimental spectra. However, we need to refine the present
model by including the crystal distortion for quantitative analysis.
We hope our simple theoretical consideration 
would stimulate further research on Sr$_2$IrO$_4$.

We are grateful to M. Yokoyama and T. Nomura for fruitful discussions. This work was partially supported by a Grant-in-Aid for Scientific Research from the Ministry of Education, Culture, Sports, Science and Technology of the Japanese Government.

\bibliography{paper}

\begin{thebibliography}{31}
\bibitem[{Com({\natexlab{a}})}]{Com1}
\bibinfo{note}{For $H_{\rm I} = \frac{1}{2}\sum_{\nu_1\nu_2\nu_3\nu_4}
  g(\nu_1\nu_2;\nu_3\nu_4) d_{\nu_1}^{\dagger}
  d_{\nu_2}^{\dagger}d_{\nu_4}d_{\nu_3}$ with $\nu=(n\sigma)$, the
  antisymmetric vertex function $\Gamma^{(0)}$ is defined as
  $\Gamma^{(0)}(\xi_{1}\xi_{2};\xi_{3}\xi_{4})
  =g(\xi_{1}\xi_{2};\xi_{3}\xi_{4})-g(\xi_{1}\xi_{2};\xi_{4}\xi_{3})$, with
  $\xi_n=(\lambda_n,\nu_n)$. It gives non-zero value only when
  $\lambda_1=\lambda_2=\lambda_3=\lambda_4$.}
\bibitem[{Com({\natexlab{b}})}]{Com2}
\bibinfo{note}{The definition of the integrated intensity of exciton peaks is
  ambiguous. We sum up intensities in the region $(\omega_0-0.2,\omega_0+0.2)$
  (eV) with $\omega_0$ being the peak position.}
\end{thebibliography}

\end{document}